\documentclass[aps,prb,twocolumn,notitlepage,superscriptaddress,showpacs]{revtex4-1}
\usepackage{graphicx,subfigure,epsfig}
\usepackage{dcolumn}
\usepackage{amssymb}
\usepackage{times}
\usepackage{amsmath}
\usepackage{amsfonts}
\usepackage{mathrsfs}
\usepackage{setspace}
\usepackage{latexsym}
\usepackage{bbm}
\usepackage{float}
\usepackage{flafter}
\usepackage{bm}
\usepackage{epstopdf}
\usepackage{hyperref}
\usepackage{color}
\usepackage{multirow}

\newcommand{\E}{\mathrm{e}}

\begin{document}

\title{Intrinsic translational symmetry breaking in a doped Mott insulator}
\author{Zheng Zhu}
\affiliation{Department of Physics, Massachusetts Institute of Technology, Cambridge, MA, 02139, USA}
\author{D. N. Sheng}
\affiliation{Department of Physics and Astronomy, California State University, Northridge, CA, 91330, USA}
\author{Zheng-Yu Weng}
\affiliation{Institute for Advanced Study, Tsinghua University, Beijing, 100084, China}
\affiliation{Collaborative Innovation Center of Quantum Matter, Tsinghua University, Beijing, 100084, China}

\date{\today}


\begin{abstract}

A central issue of Mott physics, with symmetries being fully retained in the spin background, concerns the charge excitation. In a two-leg spin ladder with spin gap, an injected hole can exhibit either a Bloch wave or a density wave by tuning the ladder anisotropy through a ``quantum critical point'' (QCP). The nature of such a QCP has been a subject of recent studies by density matrix renormalization group (DMRG). In this paper, we reexamine the ground state of the one doped hole, and show that a two-component structure is present in the density wave regime in contrast to the single component in the Bloch wave regime. In the former, the density wave itself is still contributed by a standing-wave-like component characterized by a quasiparticle spectral weight $Z$ in a finite-size system. But there is an additional charge incoherent component emerging, which intrinsically breaks the translational symmetry associated with the density wave. The partial momentum is carried away by neutral spin excitations. Such an incoherent part does not manifest in the single-particle spectral function, directly probed by the angle-resolved photoemission spectroscopy (ARPES) measurement, however it is demonstrated in the momentum distribution function. The Landau's one-to-one correspondence hypothesis for a Fermi liquid breaks down here. The microscopic origin of this density wave state as an intrinsic manifestation of the doped Mott physics will be also discussed.

\end{abstract}

\maketitle

\section{Introduction}

In a non-interacting band insulator, a doped charge behaves like a Bloch wave obeying the Bloch theorem in the presence of a periodic lattice. One may ask a similar question concerning the fate of a hole injected into a strongly correlated Mott insulator with quantum spins in the background \cite{Anderson,Lee2006}. Assume that the one-hole ground state is still translationally invariant under the translational operation by a distance ${\textbf l}$,
\begin{equation}\label{T}
\hat{T}_{\textbf{l}}|\Psi_{\mathrm {BL}}({ \textbf{k}_0}) \rangle=e^{i{\textbf k}_0\cdot {\textbf l}}|\Psi_{\mathrm {BL}}({ \textbf{k}_0})  \rangle ~,
\end{equation}
where ${\textbf k}_0$ denotes the total momentum. Then the state $|\Psi_{\mathrm {BL}}({ \textbf{k}_0})\rangle$ can be uniquely specified by the quasiparticle spectral weight
\begin{equation}\label{zk}
Z_k \equiv  \left |\langle \phi_0|c^{\dagger}_{{\textbf k}\downarrow} |\Psi_{\mathrm {BL}}({ \textbf{k}}_0) \rangle \right |^2
=\delta_{{\textbf k}, {\textbf k}_0} Z_{k_0} ~,
\end{equation}
which measures the overlap with a bare hole state of momentum ${\textbf{k}}$ created by removing a $\downarrow$ spin (without loss of generality) from the half-filling ground state $|\phi_0\rangle $. The Bloch-wave state $|\Psi_{\mathrm {BL}}({ \textbf{k}_0}) \rangle$ may generally involve a ``spin-polaron'' effect  \cite{Lee2006,SCBA1,SCBA2,SCBA3,SCBA4}, which reduces $Z_{k_0}$ but still obeying Eq. (\ref{T}).
Here Landau's one-to-one correspondence principle \cite{Andersonbook} holds true as the total momentum is completely determined by $Z_k\neq 0$ at ${\textbf k}_0$.  As the basic law of quantum mechanics, once such a Bloch state is confined within a finite-size system, the momentum quantization should naturally appear in order to make the wave function vanish at the open boundaries.

The two-leg Heisenberg spin ladder has a short-range antiferromagnetic ground state with full spin  rotational and translational symmetries\cite{Dagotto1996,White1994,Gopalan1994,Dagotto1994}. Holes injected into such a spin-gapped system can serve as an excellent example to examine elementary charge excitations in Mott systems\cite{Sigrist1994,Troyer1996,White1997,Oitmaa99,Sorella02,Scalapino,Martins2000,Lee1999,Poilblanc2003,Feiguin2008,Jeckelmann1998,Ivanov1998,Brunner2001,Dagotto1992,Tsvelik2011,Balents1996,Jiang2017,Zhu2014,Hayward 1996,Schollwock2003}. Recently  a  one-hole doped system described by the $t$-$J$ model has been systematically studied by density matrix renormalization group (DMRG) method \cite{ZZ2013,ZZ2014qp,ZZ2014cm,WSK2015}.
It has been shown \cite{ZZ2014qp} that an injected hole does propagate like a simple Bloch wave in the strong rung anisotropic limit  of the ladder. The single hole is characterized by a total momentum ${\bf k}_0=(\pi,0)$ in the ground state, with charge $+e$ and a well-defined effective mass $m^*$. The momentum is well quantized if an open boundary condition (OBC) is imposed on the finite system \cite{ZZ2013,ZZ2014qp}.

\begin{figure}[tbp]
\begin{center}
\includegraphics[width=0.48\textwidth]{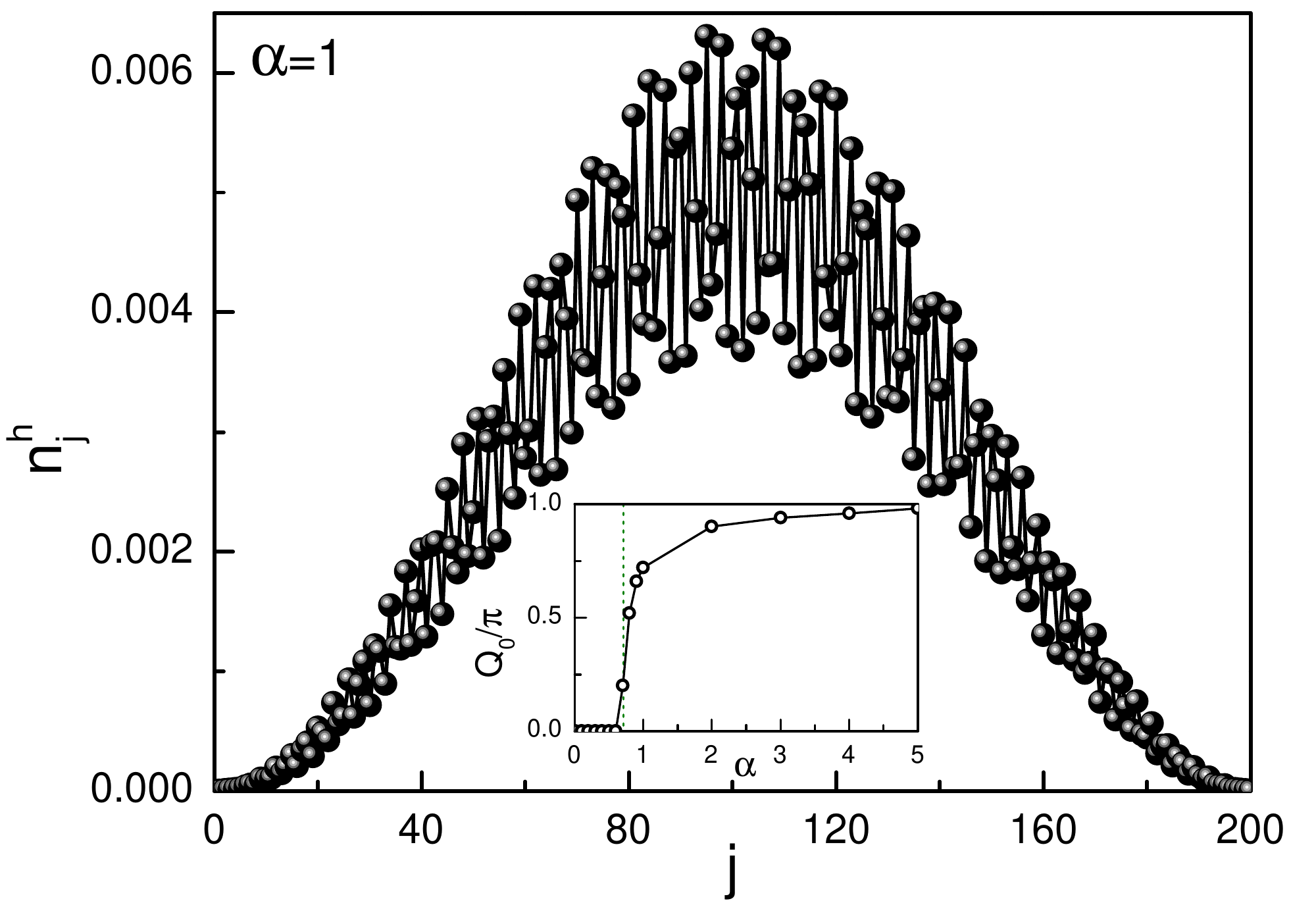}
\end{center}
\par
\renewcommand{\figurename}{Fig.}
\caption{Illustration of the density profile $n^h_j$, which exhibits a charge modulation for a single hole injected into a two-leg spin ladder governed by the $t$-$J$ model. Inset: The wave vector $Q_0$ as a function of the anisotropic parameter $\alpha$, which vanishes at $\alpha<\alpha_c$. The nature of such a one-hole state will be carefully reexamined $\alpha>\alpha_c$ by DMRG and analytic analysis in this work. (Here the ladder size $N=200\times 2$ with open boundaries, in which $j$ denoting the site along the chain direction as the density is the same on each rung, with $\alpha_c\simeq 0.7$ at $t/J=3$ based on the DMRG results \cite{ZZ2014cm}).
 }
\label{Fig0}
\end{figure}

However, a rather surprising phenomenon occurs \cite{ZZ2014qp} as the doped hole undergoes a quantum transition at a quantum critical point (QCP), $\alpha_c$, to a new state as the ladder anisotropy is continuously reduced from the strong rung limit, where the charge profile on the ladder starts to exhibit an incommensurate modulation \cite{ZZ2014cm} as illustrated in Fig. \ref{Fig0}.  This single-hole density oscillation is characterized by a nonzero wave vector ${\textbf{Q}}_0$ as a function of $\alpha$, which is the ratio of the couplings along the chain and rung directions, shown in the inset of Fig. \ref{Fig0}.  More surprisingly, the effective mass $m^*$ shows distinct behaviors depending on two \emph{different} probes after it diverges at the QCP: one ($m_s^*$) becomes finite again at $\alpha>\alpha_c$, whereas the other ($m_c^*$) remains divergent, in contrast to $m^*=m_s^*=m_c^*$ at $\alpha<\alpha_c$ \cite{ZZ2014qp}. Note that $m_c^*$ is determined by a finite-size scaling of the energy change under inserting a magnetic flux into the ring formed by the ladder \cite{ZZ2013,ZZ2014qp}. Denoting $q$ as the effective charge in units of $+e$, one has $m_c^*=m_s^*/q^2$, which thus means $q\rightarrow 0$ such that the doped hole behaves like a charge neutral ``spinon'' at $\alpha>\alpha_c$. Previously the novel property of $m_c^*\rightarrow \infty$ or $q\rightarrow 0$ has been also called self-localization of the charge degree of freedom associated with the doped hole \cite{ZZ2013,ZZ2014qp,ZZ2014cm}.

On the other hand, there is no true spin-charge separation \cite{WSK2015} as the doped hole is still composed of a pure hole (empty site) and an $\uparrow$ spin forming a \emph {loosely bound pair} at $\alpha>\alpha_c$ (in contrast to a \emph {tightly bound pair} at $\alpha<\alpha_c$) on top of a short-range antiferromagnetic background \cite{ZZ2014qp,WSK2015}. In particular, the spectral weight $Z$ has been shown \cite{WSK2015} to remain finite and smooth in the whole regime of finite $\alpha$ including at $\alpha_c$. The above-mentioned charge modulation may be understood \cite{ZZ2014cm,WSK2015} by a \emph{standing wave} density profile with the total momentum ${\bf k}_0$ split into two at ${\bf k}_0^{\pm}$. Here the nonzero wave vector ${\textbf{Q}}_0$ in the inset of Fig. \ref{Fig0} has been shown \cite{ZZ2014cm} to precisely measure such a splitting in ${\bf k}_0^{\pm}$. Then an important issue arises here, namely, whether $Z\neq 0$ would lead to the conclusion \cite{WSK2015} that the doped hole should be still in a linear combination of Bloch wave states even at $\alpha>\alpha_c$, or the standard criteria of the Landau's quasiparticle break down here such that $\alpha_c$ represents a transition between a Landau quasiparticle and a new state as a precursor of non-Fermi-liquid in the one-hole limit \cite{ZZ2013,ZZ2014qp,ZZ2014cm}.

In this paper, we shall provide numerical evidence and the combined analytic/numerical analysis to demonstrate that in the one-hole-doped case, the translational symmetry is actually spontaneously broken at $\alpha>\alpha_c$ along the ladder direction in the thermodynamic limit. Consequently, the important hypothesis in the Landau's Fermi liquid theory, i.e., the one-to-one correspondence principle by which the doped hole should carry the full total momenta, is violated in the ground state. By this understanding, the aforementioned DMRG results can be reconciled consistently.

The key results can be summarized by decomposing the one-hole ground state into \emph{two} distinct components:
\begin{equation}\label{gs}
|\Psi_{G}\rangle=c_0|\Psi_{\mathrm {stand}}\rangle+c_1|\Psi_{\mathrm {inc}}\rangle ~,
\end{equation}
where the first term $|\Psi_{\mathrm {stand}}\rangle $ is standing-wave-like composed of the Bloch waves with total momenta ${\textbf k}_0^{\pm}$, i.e., $|\Psi_{\mathrm {stand}}\rangle =\frac{1}{\sqrt{2}}\left[|\Psi_{\mathrm {BL}}({\textbf{k}}^+_0)\rangle+e^{i\phi} |\Psi_{\mathrm {BL}}({ \textbf{k}}^-_0)\rangle \right] $. Here $|\Psi_{\mathrm {BL}}({ \textbf{k}}^{\pm}_0)\rangle$ denotes a Bloch state including both the single-particle ingredient measured by the quasiparticle spectral weight $Z_k\neq 0$ at ${\textbf{k}}_0^{\pm}$ as well as a conventional spin-polaron correction \cite{Lee2006,SCBA1,SCBA2,SCBA3,SCBA4} still satisfying the translational symmetry in Eq. (\ref{T}) at the same ${\textbf{k}}_0^{\pm}$.
At $\alpha<\alpha_c$, one has $|c_1|^2=0$ with a single ${\textbf k}_0^{\pm}=k_0\equiv\pi$ such that $|\Psi_{G}\rangle$ reduces to a pure Bloch wave. Then we show that at $\alpha>\alpha_c$,  $|c_1|^2\neq 0$ such that the second component  $|\Psi_{\mathrm {inc}} \rangle$ appears in Eq. (\ref{gs}), which is orthogonal to the Bloch-wave component and no longer satisfies Eq. (\ref{T}). In other words, $|\Psi_{\mathrm {inc}} \rangle$ must possess an intrinsic translational symmetry breaking (TSB). 

We first identify $Z_k\neq 0$ at two ${\textbf{k}}_0^{\pm}$ at $\alpha>\alpha_c$ to give rise to the standing-wave component $|\Psi_{\mathrm {stand}}\rangle $ for a finite-size system. Then we further examine the spatial density profile $n^h_j$ and the momentum distribution $n^h({\textbf{k}})$ of the doped hole. A finite-size scaling analysis is used to extrapolate the DMRG results to the large sample size limit.  After subtracting the distinct behavior contributed by $|\Psi_{\mathrm {stand}}\rangle $, we can clearly identify a finite contribution from $|\Psi_{\mathrm {inc}}\rangle$ with $|c_1|^2 \neq 0$, which shows \emph{no trace} of ${\textbf{k}}_0^{\pm}$ specifying the translational invariance in Eq. (\ref{T}). Namely, an intrinsic TSB component must be simultaneously present in the ground state as $|\Psi_{\mathrm {inc}}\rangle $, which involves the charge ``incoherence'' with a continuum distribution of the momentum for the doped hole. The Landau's one-to-one correspondence is invalid here as  $|\Psi_{\mathrm {inc}}\rangle $ does not carry the same total momentum ${\textbf{k}}_0^{\pm}$ as in the Bloch component, where a partial momentum should be carried away by the background spin excitations\cite{WZ2018}. However, since such spin excitations do not directly manifest in $n^h({\textbf{k}})$ due to the Mott physics, $|\Psi_{\mathrm {inc}}\rangle $ cannot be directly probed by the single-particle spectral function $A({\textbf k},\omega)$ according to the definition, which is the physical quantity measured by the ARPES experiment. In other words, the weight of the incoherent component $|\Psi_{\mathrm {inc}}\rangle $ is totally ``missing'' from the ``Fermi surface'' probed by a conventional ARPES.

Since $|\Psi_{\mathrm {stand}}\rangle$ in Eq. (\ref{gs}) is no longer an eigenstate by itself, there must be intrinsic couplings (scatterings) between all components to lock them and minimize the total ground energy. That is, the coefficient $c_0$ and the relative phase $\phi$ inside $|\Psi_{\mathrm {stand}}\rangle $ have to be \emph{fixed} in the presence of $|\Psi_{\mathrm {inc}}\rangle $. Especially the charge oscillation is superimposed on a relatively ``flat'' background set by $|c_1|^2\neq 0$ in consistence with Fig. \ref{Fig0}, which is in sharp contrast to a pure density oscillation between the \emph{peaks and nodes} as implied by a true standing wave of the Bloch-wave states. Thus the TSB is further present in the ``standing-wave-like'' component $|\Psi_{\mathrm {stand}}\rangle $, albeit in a subtler way, by a finite-size analysis. Upon a careful examination, the ``standing wave'' component in Eq. (\ref{gs}) itself is indeed intrinsically broadened in momentum, which reveals another long but finite length scale $\lambda$. In particular, the momentum quantization of ${\textbf{k}}_0^{\pm}$ as determined by $Z_k$ fails at $\alpha>\alpha_c$, which can be also attributed to the TSB. By contrast, $Z=\sum_{\textbf k}Z_k\neq 0$ is still well converged, consistent with the previous result \cite{WSK2015}.

Finally, we discuss the microscopic origin of the intrinsic TSB  due to the precise sign structure of the $t$-$J$ model, i.e., the phase string effect. By inserting a magnetic flux in the geometry of a closed loop of the ladder along the chain direction,  the ground state energy change can be exactly formulated. A combination of the analytic analysis and DMRG calculation shows that it is the phase string as a fluctuating internal $Z_2$ field that directly responsible for ${\textbf{k}}_0^{\pm}\neq 0$ split by the incommensurate ${\textbf Q}_0$, the TSB, and charge incoherence with $q\rightarrow 0$ at $\alpha>\alpha_c$, in contrast to the diminishing phase string effect at  $\alpha<\alpha_c$ where the  the Landau's quasiparticle picture is restored.

The rest of the  paper will be arranged as follows. In Sec. II, by DMRG simulation, the quasiparticle spectral weight $Z_k$, the quantization of the total momentum and its failure at $\alpha>\alpha_c$, the momentum distribution $n^h({\textbf{k}})$, and the charge distribution along the ladder, $n^h_j$, will be presented. As an example, we shall focus on $\alpha=5>\alpha_c$ to illustrate why the translational symmetry is broken for the charge at $\alpha>\alpha_c$, and in particular, the ground state must have an incoherent part  $|\Psi_{\mathrm {inc}}\rangle $ with TSB. Then in Sec. III,
an analytic relation will be established and analyzed, which connects the origin of  $|\Psi_{\mathrm {inc}}\rangle $ and its novel properties to the phase string sign structure in the $t$-$J$ model. Finally, Sec. IV will be devoted to discussion.

\section{Translational symmetry breaking in the charge degree of freedom}

\subsection{The model }

We revisit the one-hole ground state based on the two-leg $t$-$J$ type of Hamiltonian $H_{t\text{-}J} = H_t+H_J$, in which
\begin{equation}\label{t-J}
\begin{split}
H_t &= - \sum_{\langle {ij}\rangle \sigma } t_{ij}{({c_{i\sigma }^{\dag}c_{j\sigma }+h.c.})}, \\
H_J &= \sum_{\langle {ij}\rangle } J_{ij}{(\mathbf{S}_{i}\cdot \mathbf{S}_{j}-\frac{1}{4}n_{i}n_{j})}.
\end{split}
\end{equation}
Here, $\langle ij\rangle $ stands for the nearest neighbors, ${c_{i\sigma }^{\dagger }}$ is the electron creation operator, ${\mathbf{S}_{i}}$ and ${n_{i}}$ are the spin operator and number operator, respectively, with the Hilbert space restricted by the no-double-occupancy condition $n_{i}\leq 1$. In particular, we shall study a two-leg ladder composed of two one-dimensional chains  (each with the hopping integral $t_{ij}=\alpha t$ and the superexchange coupling $J_{ij}=\alpha J$), which are coupled together by the hopping $t_{\perp}=t$ and superexchange $J_{\perp}=J$ at each rung to form a two-leg ladder \cite{ZZ2013,ZZ2014qp}. The anisotropic parameter $\alpha\rightarrow 0$ in the strong rung limit, while two chains are decoupled at $\alpha \rightarrow \infty$. We shall simultaneously study a slightly modified model known as the $\sigma$$\cdot$$t$-$J$ model \cite{ZZ2013,ZZ2014qp} for comparison, which differs from the $t$-$J$ model only by a sign factor $\sigma=\pm 1$ in the hopping term
\begin{equation}\label{sigmat}
H_{\mathrm{\sigma} \cdot t}= - \sum_{\langle {ij}\rangle \sigma } t_{ij}\sigma{({c_{i\sigma }^{\dag}c_{j\sigma }+h.c.})}~.
\end{equation}
For both models, we fix $t/J=3$ as the same in Refs.~\onlinecite{ZZ2013,ZZ2014qp,ZZ2014cm}, with the QCP $\alpha_c\simeq 0.7$ \cite{ZZ2014qp,ZZ2014cm,WSK2015}.

In the following, the one-hole ground state (with a down-spin electron removed from the half-filling) of the two-leg ladder (of the sample size $N= N_x\times 2$) is computed by using the DMRG algorithm. In the calculations, we keep up to around $1800$ states, which controls the truncation error to be in the order of $10^{-10}$ and $10^{-6}$ for open and periodic systems, respectively.
In the calculation of the spectral weight $Z_k$ or $Z_j$, we do more than 200 sweeps to obtain well converged results.

\begin{figure*}[tbp]
\begin{center}
\includegraphics[height=4.4in,width=6in]{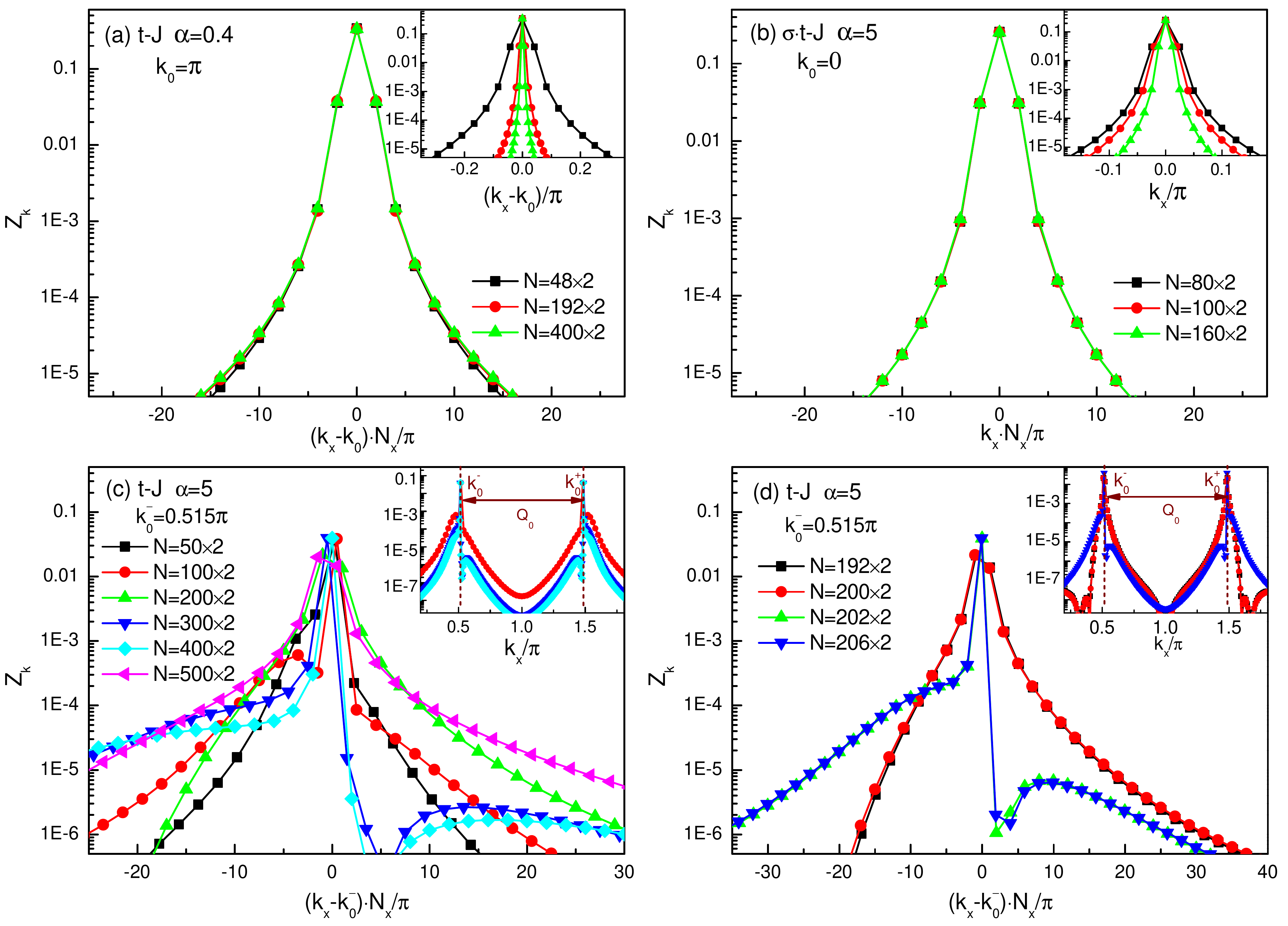}
\end{center}
\par
\renewcommand{\figurename}{Fig.}
\caption{(Color online) The quasiparticle spectral weight $Z_k$ can determine the total momentum ${\bf k}_0\equiv (k_0, k_{0y})$ via a finite-size analysis. Insets: the original $Z_k$'s. (a) A well-quantized Bloch wave, with $k_0=\pi$ and $k_{0y}=0$ in the $t$-$J$ case at $\alpha=0.4<\alpha_c$, is characterized by the scaling law with the $k_x$-axis replaced by $(k_x-k_0)N_x/\pi$; (b) A well-quantized Bloch wave with $k_0=0$ in the $\sigma$$\cdot$$t$-$J$ ladder at $\alpha=5$; (c) and (d) The $t$-$J$ case at $\alpha=5>\alpha_c$: the quantization in a finite-size sample breaks down even at small variations of the sample length, for example, $N_x=192$, $200$, $202$, and $206$ [cf. (d)]. Here the total momentum $k_0$ is split into two $k^{\pm}_0$ separated by an incommensurate $Q_0$ defined by Eq. (\ref{Q0}) with $k_{0y}=0$.
}
\label{Fig:Zk}
\end{figure*}

\subsection{Quasiparticle spectral weight}

In general, the quasiparticle spectral weight $Z_k$ is defined to measure the probability that the one-hole ground state $|\Psi_G \rangle$ is projected onto a bare Bloch wave state
\begin{equation}\label{Bloch}
|{\textbf{k}}\rangle=\sqrt{2} c_{{\textbf k}\downarrow}|\phi_0\rangle ~,
\end{equation}
by
\begin{equation}\label{zk1}
Z_k\equiv \left |\langle \phi_0|c^{\dagger}_{\textbf{k}}|\Psi_G \rangle \right |^2 =
\frac 1 2 \left |\langle {\textbf{k}}|\Psi_G \rangle \right |^2
\end{equation}
If both states obey the translational symmetry, $Z_{k}$ may be utilized to determine the total momentum ${ \textbf{k}}_0$ of $|\Psi_G \rangle$ as pointed out in the Introduction [cf. Eq. (\ref{zk})]. Note that even in the one-dimensional (1D) chain with one hole, where $Z_{k_0}$ eventually vanishes  as $\propto (N)^{-\gamma}$ at $N \rightarrow \infty$ (i.e., a Luttinger liquid behavior), one still can have $Z_{k}\neq 0$ sharply peaked around $k_0=\pm \pi/2$ at a finite but large $N$ \cite{Zhu2016}.

But $Z_{k_0}$ cannot directly measure the second term in the ground state Eq. (\ref{gs}) as $\langle {\textbf{k}}|\Psi_{\mathrm {inc}}\rangle = 0$. In other words, for $Z_k\neq 0$ to signal the existence of a coherent quasiparticle, there should be an underlying assumption that $|\Psi_{\mathrm {inc}}\rangle=0 $ without TSB. This is actually the famous one-to-one correspondence hypothesis \cite{Andersonbook} of the Landau's Fermi liquid theory.
Only in this case can a finite $Z_k$ fully characterize the doped hole as a quasiparticle excitation.
Nevertheless, in the following, we shall show that even if $|\Psi_{\mathrm {inc}}\rangle $ with TSB appearing in the one-hole ground state Eq. (\ref{gs}) with $|c_1|^2\neq 0$, $Z_k$ can still provide an important and distinct signal in the finite-size analysis.

\subsubsection{The quantization of total momentum }

We first inspect $Z_k$ at $\alpha=0.4<\alpha_c$. As pointed out above, $Z_k$ can directly determine the total momentum in a translational invariant system. As shown in Fig.~\ref{Fig:Zk}(a), $Z_k$ is found to be peaked at ${\textbf{k}}_0=(k_0, k_{0y})$ with $k_0=\pi$ and $k_{0y}=0$ for the $t$-$J$ model. With an OBC, the translational symmetry is slightly broken such that a small range of momenta around ${\textbf{k}}_0$ is involved.
 Then the wave quantization should be seen in a finite size scaling for finite size systems.  Indeed, the data presented in the inset of Fig.~\ref{Fig:Zk}(a) can be well collapsed under a rescaling
 \begin{equation}\label{scalingI}
k_x \rightarrow (k_x-{k}_0)N_x
\end{equation}
in the main panel. They clearly indicate that the doped hole behaves like a coherent Bloch wave in the large $N_x$ limit, where the ground state converges to a single momentum ${ \textbf{k}}_0$. Figure~\ref{Fig:Zk}(b) shows a similar Bloch wave behavior for the single hole in the $\sigma$$\cdot$$t$-$J$ model with $ k_0=0$ at a much larger $\alpha=5$ (to be compared with the $t$-$J$ case below).

Now let us focus on $\alpha>\alpha_c$ for the $t$-$J$ model. An emerging double-peak structure centered at $k_0^{\pm}$ is shown in the inset of Fig.~\ref{Fig:Zk}(c) at $\alpha=5>\alpha_c$.  Here the total momentum ${\textbf{k}}_0=(k_0, 0)$ with $k_0$ starting to split by $Q_0$ as
\begin{equation}\label{Q0}
k_0^{+}-k_0^{-}\equiv Q_0
\end{equation}
for the $t$-$J$ model, which is consistent with $Q_0$ previously determined by different methods \cite{ZZ2014qp,ZZ2014cm,WSK2015} [cf. the inset of Fig. \ref{Fig0}(a)].

Here a new feature besides the momentum splitting along $k_x$ is that the \emph{wave quantization} under the OBC is no longer valid, as clearly shown in the main panel of Fig.~\ref{Fig:Zk}(c), even though $N_x$ is much larger than the size of the quasiparticle presumably decided by the spin-spin correlation length \cite{ZZ2014qp}.
As a matter of fact, the distribution of momenta strongly scatter around $k_0^{\pm}$ even under some very small changes of sample sizes, e.g., comparing $N_x=192$, $200$, $202$, and $206$ in Fig.~\ref{Fig:Zk}(d).

Therefore, at $\alpha>\alpha_c$, although the weight of $Z_k$ is converged to $k_x=k_0^{\pm}$ in the limit of $N_x\rightarrow \infty$, the finite-size scaling associated with the momentum quantization is absent. It is well known in the quantum mechanics that the momentum quantization is a basic signature for a free wave confined in a box. Its absence on the  $\alpha>\alpha_c$ side indicates that the single-hole state may have gained a nontrivial many-body component $|\Psi_{\mathrm {inc}}\rangle $, which cannot be reduced to a Bloch wave in Eq. (\ref{gs}). In this case, when one imposes the OBC onto the wave function, the quantization of the ``standing wave'' component will get scrambled as $|c_1|^2\neq 0$.

\begin{figure}[tbp]
\begin{center}
\includegraphics[width=0.48\textwidth]{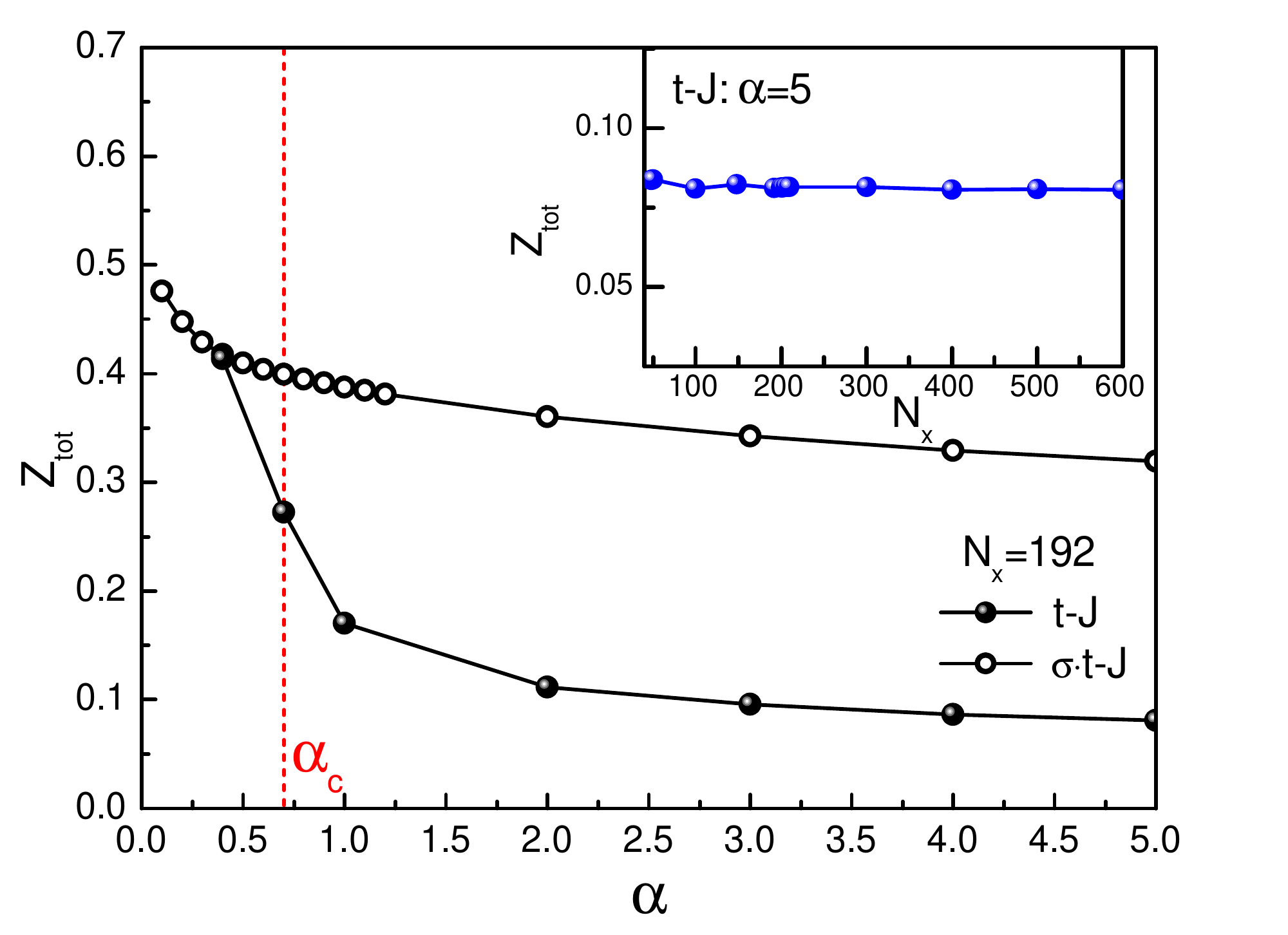}
\end{center}
\par
\renewcommand{\figurename}{Fig.}
\caption{(Color online) $Z_{\mathrm{tot}}$ measures the overlap of the true ground state with a bare hole state, but it does not solely decide the coherence of the quasiparticle should the one-to-one correspondence principle fails (see text). Here $\alpha$ denotes an anisotropic parameter for the two-leg $t$-$J$ ladder. A critical point $\alpha_c$ is marked by the vertical dashed line, which is previously determined \cite{ZZ2014qp} for the $t$-$J$ case at $t/J=3$ with $\alpha_c\approx 0.7$ (no critical $\alpha_c$ for the $\sigma$$\cdot$$t$-$J$ model). Inset: the convergence of $Z_{\mathrm{tot}}$ vs. the sample size $N= N_x\times 2$ at $\alpha=5$.  }
\label{Fig1}
\end{figure}


\subsubsection{The total single hole spectral weight}

Without the momentum quantization at $\alpha>\alpha_c$, $Z_k$ itself may not be a good measure of the bare hole weight. Instead, one may introduce the \emph{total} spectral weight satisfying
\begin{equation}\label{ztot}
Z_{\mathrm {tot}}\equiv \sum_k Z_k= \sum_j Z_j~,
\end{equation}
where $Z_j\equiv \left |\langle \phi_0|c^{\dagger}_{j\downarrow} |\Psi_G \rangle \right |^2=1/2\left |\langle j|\Psi_G \rangle \right |^2$ is proportional to the probability of the ground state $|\Psi_G \rangle$ projected onto $|j\rangle\equiv \sqrt{2}c_{j\downarrow} |\phi_0 \rangle $. Hence, $Z_{\mathrm {tot}}$ characterizes the total weight of a \emph{bare} hole state in the true one-hole ground state. In particular, for a Bloch wave state of momentum ${\textbf{k}}_0$, $Z_k$ is given by Eq. (\ref{zk}).

A finite $Z_{\mathrm {tot}}$ for the $t$-$J$ ladder is computed by DMRG as shown in Fig.~\ref{Fig1} (full circle). A single doped hole should only change the spin background around it, independent of a sufficiently large $N_x$, since the spin excitation is always gapped in the two-leg spin ladder at half-filling. In the inset of Fig.~\ref{Fig1}, $Z_{\mathrm {tot}} $ is quickly saturated with the increase of $N_x$, which is in sharp contrast to the scattering data of $Z_k$ illustrated in Figs.~\ref{Fig:Zk}(c) and (d). $Z_{\mathrm {tot}} $ remains finite across the QCP $\alpha_c$, marked by a vertical dashed line previously determined in Ref. \onlinecite{ZZ2014qp} [cf. the inset of Fig. \ref{Fig0}(a)].
For comparison, $Z_{\mathrm {tot}}$ for the aforementioned $\sigma$$\cdot$$t$-$J$ ladder is also presented in Fig.~\ref{Fig1} (open circle). Note that for such a model, there is no critical point throughout the whole $\alpha$ regime \cite{ZZ2014qp}.

$Z_{\mathrm {tot}}$ in Fig.~\ref{Fig1} is in good agreement with $Z$ obtained \cite{WSK2015} by a slightly different method for the $t$-$J$ case (there is a factor 2 difference is due to a normalization factor in defining the bare hole state, e.g., at $\alpha=1$, $Z_{\mathrm {tot}}=0.17037$ as compared to $0.34067$ in Ref. \onlinecite{WSK2015}). However, a finite $Z$ has been used \cite{WSK2015} as one of the key evidence to support the argument that the doped hole should always behave like a Bloch quasiparticle, and $\alpha_c$ would simply separate two regimes of Bloch wave states differed only by being non-degenerate and doubly degenerate, respectively. On the other hand, the emergence of the incoherent part $|\Psi_{\mathrm {inc}}\rangle $ with TSB will mean that even a finite $Z_{\mathrm {tot}}$ is no longer sufficient to imply the existence of a coherent Bloch quasiparticle at $\alpha>\alpha_c$ as to be discussed below.

\begin{figure*}[tbp]
\begin{center}
\includegraphics[height=3.5in,width=7.0in]{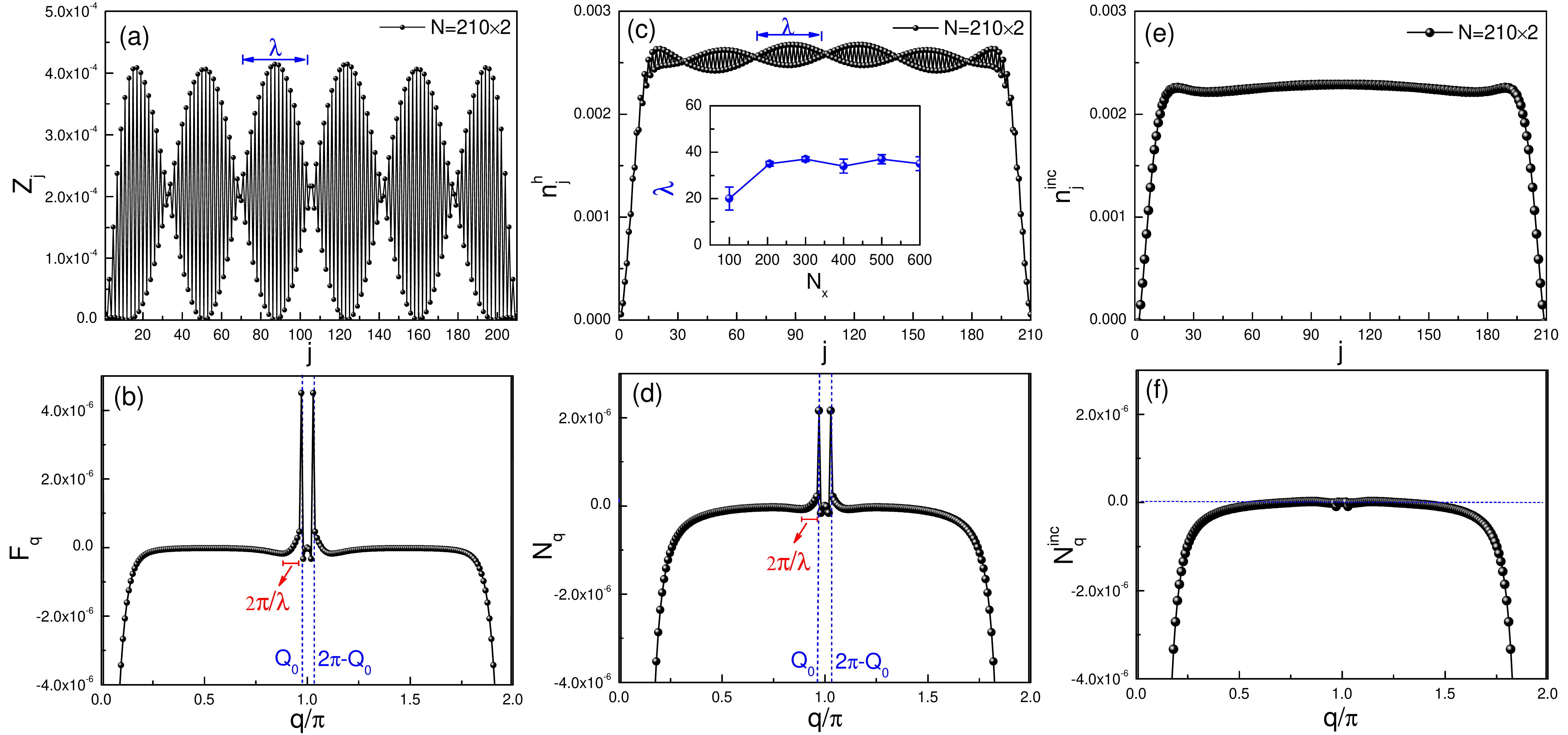}
\end{center}
\par
\renewcommand{\figurename}{Fig.}
\caption{(Color online) The origin of charge density modulation in $n_j^h$ can be fully attributed to that of the standing-wave component in the ground state Eq. (\ref{gs}), whose probability is measured by $Z_j$. (a):  A fast oscillation in  $Z_j$ is modulated by a slower variation at a length scale of $\lambda$ at $\alpha=5>\alpha_c$; (b): The Fourier transformation of $Z_j$, $F_q$, reveals the characteristic wave vector $Q_0$ with a continuous spread $\sim 2\pi/\lambda$; (c) and (d): The corresponding hole density distribution $n_j^h$ and its Fourier transformation. Inset of (c): The length scale $\lambda$ vs. $N_x$. Finally, (e) and (f):  the contribution from $|\Psi_{\mathrm {inc}}\rangle $ estimated by $n_j^{\mathrm {inc}}\equiv n_j^h-Z_j-(\beta Z_j+Z_{\text{tot}}/N)$ ($\beta\simeq -1/2$) and its Fourier transformation $N_q^{\mathrm {inc}}=N_q-(1+\beta)F_q$, which show negligible traces of the wave vector $Q_0$ (see text).}
\label{Fig3}
\end{figure*}

\subsection{Origin of charge modulation}

It has been previously found \cite{ZZ2014cm} that there is always a charge density modulation characterized by the wave vector $Q_0$ at $\alpha>\alpha_c$ (cf. the inset of Fig. \ref{Fig0}). An important issue is whether it is a robust phenomenon associated with the TSB in the charge degree of freedom, or, in an alternative view, is simply the manifestation of a standing wave composed of two degenerate Bloch states \cite{WSK2015}. In the latter scenario, the charge modulation would be merely an artifact of the double degeneracy of the ground states, which may be easily lifted without intrinsic protection.

Let us first identify the origin of this charge density modulation by examining the real space distribution of $Z_j$, the probability of a bare hole state $|j\rangle$ in the one-hole ground state $|\Psi_G \rangle$ as previously defined in Sec. II B2. Based on Eq. (\ref{gs}), one finds that the contribution of $|\Psi_{\mathrm {stand}}\rangle $ gives rise to
\begin{align}\label{zj}
Z_j &=\frac{Z_{\text{tot}}}{N}\left[1+ \cos({\textbf Q}_0\cdot {\bf r}_j+\phi)\right]  ~,
\end{align}
with $Z_{\text{tot}}\propto |c_0|^2$ and $\phi$ as a relative phase. The calculated $Z_j$ by DMRG and its Fourier transformation $F_q$ at $\alpha=5$ are presented in Figs.~\ref{Fig3}(a) and \ref{Fig3}(b), respectively. One sees that a sharp spatial oscillation characterized by $Q_0$ is indeed present, which matches with the momentum splitting by Eq. (\ref{Q0}) (note that there is an additional slow modulation of a longer length scale $\lambda$, which is to be discussed below).

On the other hand, the hole density distribution $n_j^h$ and its Fourier transformation $N_q$ at finite wave vector $q$ part are shown in Figs.~\ref{Fig3}(c) and \ref{Fig3}(d), respectively.  Indeed the density oscillation in Figs.~\ref{Fig3}(c) and the finite wave vector peaked at $Q_0$ in \ref{Fig3}(d) match with the features of $Z_j$ and its Fourier transformation in Figs.~\ref{Fig3}(a) and \ref{Fig3}(b). The translational symmetry makes $|\Psi_{\mathrm {stand}}\rangle $ to be transformed as a linear combination of Eq. (\ref{T}) with momenta ${\textbf k}^{\pm}_0$, which results in the density oscillation of the wave vector $Q_0$.  Thus the density oscillation in $n_j^h$ should be \emph{fully} accounted for by the standing-wave component in Eq. (\ref{gs}), which is constituted by the quasiparticle component $Z_j$ and a portion $\beta Z_j$ due to the spin-polaron effect. We find $\beta\simeq -1/2$ at $\alpha=5$ according to our DMRG fitting such that the total contribution from the standing-wave oscillation should be $Z_j+(\beta Z_j+ Z_{\text{tot}}/N)$ where in the second (i.e., the spin-polaron correction) term a constant is added to make it non-negative. Then, the contribution from $|\Psi_{\mathrm {inc}}\rangle $ can be deduced as the flat background given in Fig.~\ref{Fig3}(e) and its Fourier transformation in Fig. \ref{Fig3}(f) with $n_j^h$ being subtracted by the aforementioned standing-wave contribution. The latter as due to the TSB component $|\Psi_{\mathrm {inc}}\rangle $ shows \emph{no oscillation} or the trace of $Q_0$, with $|c_1|^2$ estimated by $1-(3/2)Z_{\text{tot}}\simeq 0.88$ at $\alpha=5$.

Furthermore, it is important to observe that the Bloch-type standing-wave component, $|\Psi_{\mathrm {stand}}\rangle$, which gives rise to a finite $Z_k$ or $Z_{\mathrm {tot}}$, is \emph{no longer} an independent eigenstate, but only an integral part of the true ground state $|\Psi_G \rangle$ in Eq. (\ref{gs}). By mixing with the incoherent component $|\Psi_{\mathrm {inc}}\rangle $, $|\Psi_{\mathrm {stand}}\rangle$ should also become TSB in general. Previously, we have already shown the non-quantization of the momenta ${\textbf k}^{\pm}_0$ in a finite-size system. Figure~\ref{Fig3}(a) further indicates another slower spatial modulation of a length scale $\lambda$, which corresponds to a continuous broadening around $Q_0$ in Fig.~\ref{Fig3}(b). The incoherent length scale $\lambda$ vs. $N_x$ is plotted in the inset of Fig.~\ref{Fig3}(c).
It implies that the ``coherent'' component $|\Psi_{\mathrm {stand}}\rangle$ in Eq. (\ref{gs}) is by itself also TSB indeed, if viewed from a sufficiently long distance ($>\lambda$).

\begin{figure}[tbp]
\begin{center}
\includegraphics[width=0.48\textwidth]{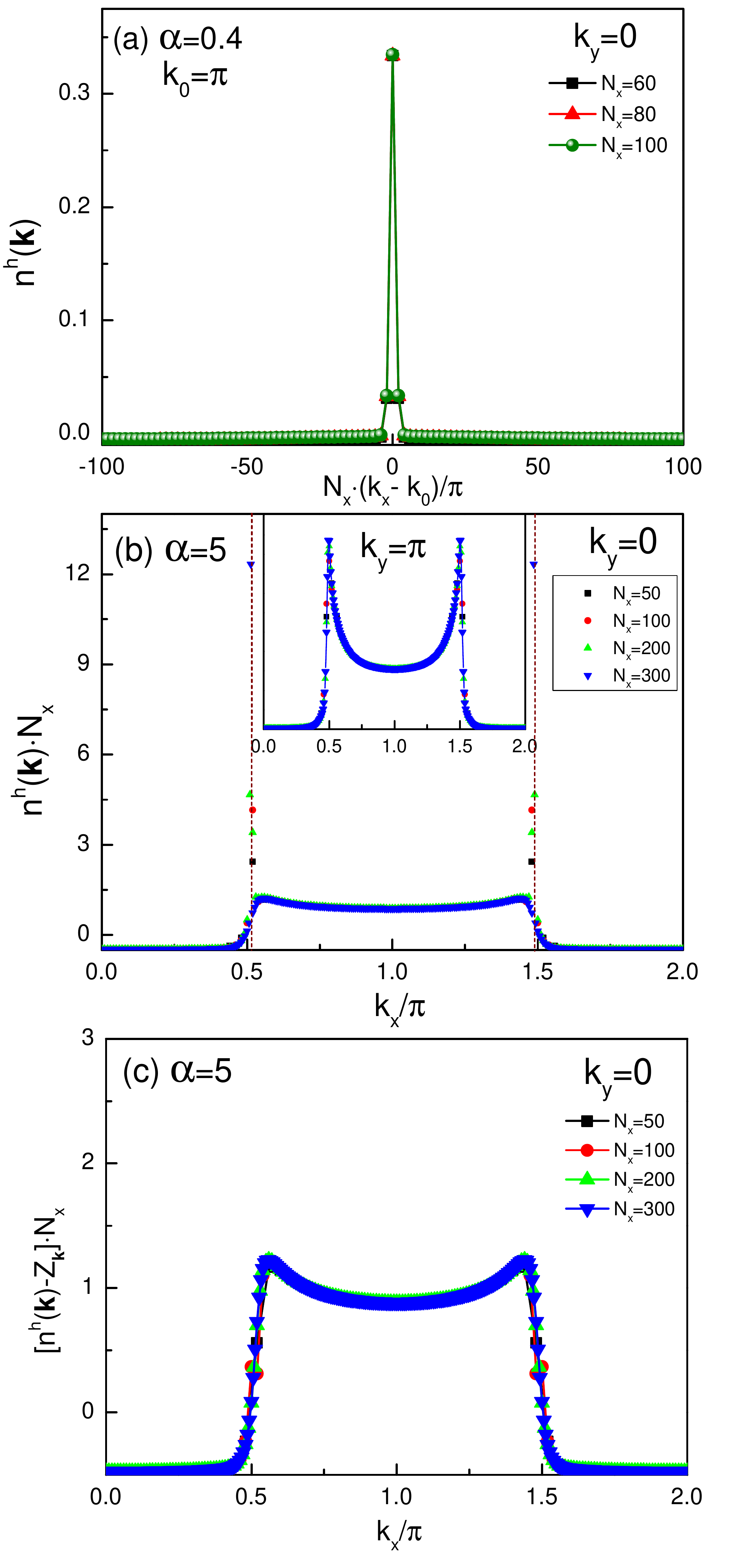}
\end{center}
\par
\renewcommand{\figurename}{Fig}
\caption{(Color online) Distinctive momentum distributions of the hole in the two phases separated by $\alpha_c=0.7$: (a) At $\alpha=0.4<\alpha_c$, $n^h({ \textbf{k}})$ shows a finite-size scaling similar to $Z_k$ in Fig. 2(a) in consistency with a Bloch wave description;
(b) At $\alpha=5>\alpha_c$, besides two sharp peaks located around the total momentum $k^{\pm}_0$ represented by $Z_k$ (marked by the vertical dashed lines), an additional continuous background (blue curves), respectively, at both $k_y=0$ and $k_y=\pi$ (the inset), satisfies a different scaling law, i.e., $n^h({ \textbf{k}})\cdot N_x $, which ensures a finite weight in the sum rule Eq.(\ref{sumr}); (c) The broad feature at $k_y=0$ in the main panel of (b) is re-drawn as $[n^h({ \textbf{k}})-Z_k]\cdot N_x $. }
\label{Fig:nk}
\end{figure}

\subsection{The breakdown of the one-to-one correspondence}

To compare with the ``total momentum'' determined by $Z_k\neq 0$ above, one may calculate the momentum distribution of the doped hole, defined by
\begin{equation}\label{nk}
n^h({\textbf{k}})\equiv 1-\sum_{\sigma}\langle \Psi_G|c^{\dagger}_{ \textbf{k} \sigma}c_{\textbf{k}\sigma}|\Psi_G \rangle ~,
\end{equation}
which satisfies the following sum rule
\begin{equation}\label{sumr}
\sum_{\textbf{k}} n^h({\textbf{k}})=1
\end{equation}
as there is only one hole with the total momentum given by
\begin{equation}\label{k0}
{\textbf{k}}_0^h=\sum_{\textbf{k}} {\textbf{k}} n^h({\textbf{k}})~.
\end{equation}
Here ${\textbf{k}}_0^h$ in general can be different from the total momentum should the neutral spin background acquires finite momentum excitations. Note that $n^h({\textbf{k}})\equiv 0$ at half-filing even if there is a spin excitation carrying a finite momentum such that $n^h({\textbf{k}})$ can \emph{only} measure the momentum associated with the doped charge (hole). This is a peculiar property of the Mott insulator in which the electrons are all localized onsite at half-filling. In other words, $n^h({\textbf{k}})\neq 0 $ will determine the Fermi surface (points) for the present one-hole state, but could represent a ``wrong'' total momentum structure should the novel spin excitation is also present. This will be the issue to be carefully examined below.

The peak of $n^h({\textbf{k}})$ does coincide with the total momentum ${\textbf{k}}_0=(\pi,0)$ at $\alpha=0.4<\alpha_c$, as shown in Fig.~\ref{Fig:nk}(a). Note that the whole data are well collapsed by the rescaling in Eq. (\ref{scalingI}) along the $k_x$-axis. The finite-size scaling analysis \cite{ZZ2013,ZZ2014qp,ZZ2014cm} suggests that the momentum ${\textbf k}$ approaches ${\textbf k}_0$ at $N_x\rightarrow \infty$, which is consistent with that of $Z_k$ in Fig. \ref{Fig:Zk}(a). Thus, the one-to-one correspondence principle is still valid, and the doped hole carries the \emph{total} momentum as a Bloch quasiparticle, with the full ground state obeying the translational symmetry Eq. (\ref{T}).

As shown in Fig.~\ref{Fig:nk}(b) at $\alpha=5>\alpha_c$, $n^h({\textbf{k}})$ is still peaked at the same positions of the momenta ${\textbf{k}}_0\equiv (k_0^{\pm}, 0)$ as identified by $Z_k$ in Fig. \ref{Fig:Zk}(c). However, $n^h({\textbf{k}})$ also gains an additional continuous background, which no longer satisfies the scaling behavior in Eq. (\ref{scalingI}). In contrast,  a new scaling behavior is found for the continuous backgrounds of the momentum in the main panel of Fig.~\ref{Fig:nk}(b) ($k_y=0$) and the inset ($k_y=\pi$). Namely, one finds that $n^h({\textbf{k}})-Z_k$ will collapse onto a universal curve$/N_x$, which measures the contribution from $|\Psi_{\mathrm {inc}}\rangle $, persisting in the thermodynamic limit to make a \emph{finite} contribution in the sum rule of Eq. (\ref{sumr}). Here $[n^h({\textbf{k}})-Z_k]\cdot N_x$ at $k_y=0$ is re-drawn in Fig.~\ref{Fig:nk}(c). On the other hand, the single-particle contribution of the standing wave component $|\Psi_{\mathrm {stand}}\rangle$ in Eq. (\ref{gs}) is represented by $Z_k$ [sharp peaks indicated in the main panel of Fig.~\ref{Fig:nk}(b)], which does not satisfy this scaling law.
At $\alpha=5$, we find $Z_{\mathrm{tot}}=\sum_k Z_k \sim 0.08$, while the broad background in the main panel of Fig.~\ref{Fig:nk}(b) gives rise to a total weight of $\sim 0.26$ at $k_y=0$.

The rest of contribution, about $0.66$, will come from $k_y=\pi$ as shown in the inset of
Fig.~\ref{Fig:nk}(b), where a broad double-peak structure satisfies the same scaling behavior along $k_x$. In other words, in the thermodynamic limit, a finite portion of the hole will carry a continuum of momenta at $k_y=\pi$ where $Z_k$ remains exponentially small [not shown in Fig. \ref{Fig:Zk}(c)].

Therefore, the hole has a continuous distribution of momentum as illustrated in Fig.~\ref{Fig:nk}(b) as given by $n^h({\textbf{k}})-Z_k$, which accounts for a weight approximately $\sim 0.92$ at $\alpha=5$. This is quite substantial in comparison to the two isolated momenta at ${\textbf{k}}^{\pm}_0=(k_0^{\pm},0)$ as indicated by $Z_k\neq 0$. We note that even though such a broad feature may also include the contribution from the spin-polaron correction with $Z=0$ and the well defined total momenta ${\textbf{k}}^{\pm}_0$,   when the spin-polaron correction in $|\Psi_{\mathrm {stand}}\rangle$ is considered as in the previous subsection, the total weight of the incoherent $|\Psi_{\mathrm {inc}}\rangle $ is still around a substantial $|c_1|^2 \simeq 0.88$. Hence, instead of a ``marginal Fermi liquid'' behavior \cite{Anderson} via $Z_{k} \rightarrow 0$, a new way approaching a non-Fermi-liquid has been identified here by violating the one-to-one correspondence hypothesis with an incoherent $|\Psi_{\mathrm {inc}}\rangle $ with TSB in Eq. (\ref{gs}).

However, since the incoherent $|\Psi_{\mathrm {inc}}\rangle $ has zero overlap with the bare Bloch state $c_{{\textbf{k}}\downarrow}|\phi_0 \rangle$, it cannot be probed by the spectral function $A({ \textbf{k}},\omega)$.  In a conventional Green's function or the  time-dependent DMRG (tDMRG) approach \cite{Andersonbook,WSK2015}, $A({ \textbf{k}},\omega)$ is given by
\begin{align}\label{A}
A({\textbf{k}},\omega)&=2\pi \sum_{n }\left |\langle \Psi_n |c_{{\textbf{k}}\downarrow}|\phi_0 \rangle \right |^2\delta(\omega-E_{n0}(\textbf{k})) \nonumber \\
&\equiv 2\pi Z_k \delta(\omega)+ A_{\mathrm{cont}}({\textbf{k}},\omega)~,
\end{align}
by following the time-evolution of a bare Bloch hole created on $|\phi_0\rangle $, where $|\Psi_n \rangle$ denotes a one-hole eigenstate with energy $E_{n0}(\textbf{k})$. So $A({\textbf{k}},\omega)$ measures the probability of the bare Bloch state $c_{{ \textbf{k}}\downarrow}|\phi_0 \rangle$ in one-hole eigenstates at energy $\omega$, which decides the quasiparticle spectral weight $Z_k$ at $\omega=0$ (i.e., in the ground state $|\Psi_0 \rangle\equiv |\Psi_G \rangle$). The tDMRG result shows \cite{WSK2015} that at $\omega=0$, only $k^{\pm}_0$ contribute to the poles, which are separated by an order of $J$ from the continuum $A_{\mathrm{cont}}({\textbf{k}},\omega)$ that involves all the excited ($n\neq 0$) $|\Psi_n \rangle$. Note that it should not be confused with the ``incoherent part'' $|\Psi_{\mathrm {inc}}\rangle $ in the ground state, as the former is originated from the excited states at \emph{higher energies}.

So we see that the usual criterion for identifying a Landau quasiparticle by $Z_k$ or $Z_{\mathrm {tot}} \neq 0$ is no longer applicable if the one-to-one correspondence breaks down. In the present one-hole ground state, $n^h({\textbf{k}})$ clearly shows that besides $k^{\pm}_0$, there is also a significant incoherent weight contributed by $|\Psi_{\mathrm {inc}}\rangle $, but not detectable by $A({\textbf{k}},\omega)$ by definition. As such, one has to be very careful in utilizing the conventional Green's function or tDMRG analysis to identify the ``coherent quasiparticle'', even as $Z_{\mathrm {tot}} \neq 0$, in a strongly correlated system. The ``dark matter'' represented by $|\Psi_{\mathrm {inc}}\rangle $ is intrinsically TSB involving a continuum of momentum distribution, which has been consistently identified by examining the spatial density profile in Fig.~\ref{Fig3}. Its physical origin will be discussed in Sec. III below.

\section{Microscopic origin of $Q_0$}

In the previous section, we have provided a series of DMRG evidence to support the existence of an intrinsic TSB phase at $\alpha>\alpha_c$ in the one-hole-doped ground state of the $t$-$J$ ladder. On the other hand, it has been also found \cite{ZZ2013} that such a novel phase completely disappears in the $\sigma\cdot$$t$-$J$ model with a spin-dependent hopping, which differs from the $t$-$J$ model only by a phase-string sign structure. In the following, we establish a direct analytic connection between the TSB and the microscopic description of the phase string.

\subsection{Phase-string sign structure of the  $t$-$J$ model}

Let us start by briefly reviewing the generic mathematical structure of the sign structure in the $t$-$J$ model for the sake of self-containedness.

For the one-hole-doped $t$-$J$ model, an exact expression of the partition function is given by \cite{Wu2008sign}
\begin{equation}\label{Zc}
\mathcal{Z}_{t\text{-}J}=\sum_{c}{\tau }_{c}\mathcal{W}[c]~,
\end{equation}
where the hole acquires a Berry-like phase \cite{Sheng1996} as
\begin{equation}\label{tauc}
\tau _c  = \left( { - 1} \right)^{N_h^ \downarrow  [c]}=\pm 1
\end{equation}
going through a closed path $c$ (a brevity for multi-paths of the spins and the hole). Here $N_{h}^{\downarrow }[c]$ counts the total number of exchanges between the hole and down spins. The weight $\mathcal{W}[c]\ge 0$ is dependent on temperature ($1/\beta$), $t$, $J$, and $\alpha$ \cite{Wu2008sign} and the total lattice size $N$ is assumed to be even (bipartite).

Thus, each closed path associated with the doped hole is always modulated by a unique sign factor (\ref{tauc}). It is called\cite{Sheng1996} the phase string effect and the underlying picture may be understood as follows. In general, a doped hole moves on a spin background governed by the $t$-$J$ Hamiltonian will create a string of spin displacement on its path, which is of three (i.e., spin x, y and z) components. One of the components (say, spin-z) can be always ``repaired" through spin flip process, but the other two, i.e., the transverse ones, cannot be simultaneously self-healed via the Heisenberg term and thus be left behind as a sequence of $Z_2$ signs, precisely described by $\tau_c$ as given by Eq. (\ref{tauc}).

Utilized as a comparable study, the so-called $\sigma$$\cdot$$t$-$J$ model has been introduced in Ref. \onlinecite{ZZ2013} by inserting a spin-dependent sign in the hopping term of the $t$-$J$ model [with a tight-binding hopping term $-t\sigma c_{i\sigma}^+c_{j\sigma}$ as given in Eq. (\ref{sigmat})], such that the one-hole partition function reduces to \cite{ZZ2013}
\begin{equation}\label{Zstj}
\mathcal{Z}_{\sigma\cdot t\text{-}J}=\sum_{c}\mathcal{W}[c]~,
\end{equation}
which is only different from $\mathcal{Z}_{t\text{-}J}$ [Eq.~(\ref{Zc})] by the absence of the Berry-like phase ${\tau }_{c}$, with the same $\mathcal{W}[c]$.

Next, we further point out that the quasiparticle spectral weight $Z_k$ and $Z_j$  for the t-J model are determined by the single hole propagator, which also can be formally expressed as \cite{Sheng1996,Wu2008sign}
\begin{equation}\label{G}
G_h(i, j; E)\propto \sum_{c_{ij}}\tau_{c_{ij}} P(c_{ij})
\end{equation}
in which the phase string factor $\tau_{c_{ij}}$ modulates each path $c_{ij}$, including all the paths of spins and the hole, with the hole path connecting site $i$ and $j$, with a weight $P(c_{ij})>0$ \footnote{With $E$ less than the ground state energy $E_G^{\text{1-hole}}$}. According to Eq.~(\ref{tauc}), one may show \cite{ZZ2014cm} that the momentum structure comes from the phase string factor
\begin{equation}\label{psij}
\tau_{c_{ij}} \sim \E^{i\mathbf {k}_0^{\pm}\cdot [{\mathbf r}_i-{\mathbf r}_j]+i\delta_{ij}} ~,
\end{equation}
in which $\mathbf {k}_0^{\pm} \cdot [{\mathbf r}_i-{\mathbf r}_j]$ denotes an averaged $N_h^ \downarrow(c_{ij})$ with $\mathbf {k}_0^{\pm}$ characterized by $Q_0$ in (\ref{Q0}), and the phase shift $\delta_{ij}$ captures the rest of many-body fluctuations around $\mathbf {k}_0^{\pm}$. The phase shift $\delta_{ij}$ is the source leading to the incoherence of the charge. As a matter of fact, by switching off $\tau_c$, all the modulations disappear in $Z_j$ and $n_j^h$ in the $\sigma\cdot$$t$-$J$ model \cite{ZZ2013,ZZ2014qp,ZZ2014cm}.

\subsection{Determining $Q_0$ based on the phase string effect}

In the following, we shall use the exact expression of Eq. (\ref{Zc}) to study the charge response to inserting a magnetic flux $\Phi$ into a ring of the ladder enclosed along the chain direction, by which the microscopic origin of $Q_0$ and the charge incoherence will be determined quantitatively.

\begin{figure}[t]
\begin{center}
\includegraphics[width=\columnwidth]{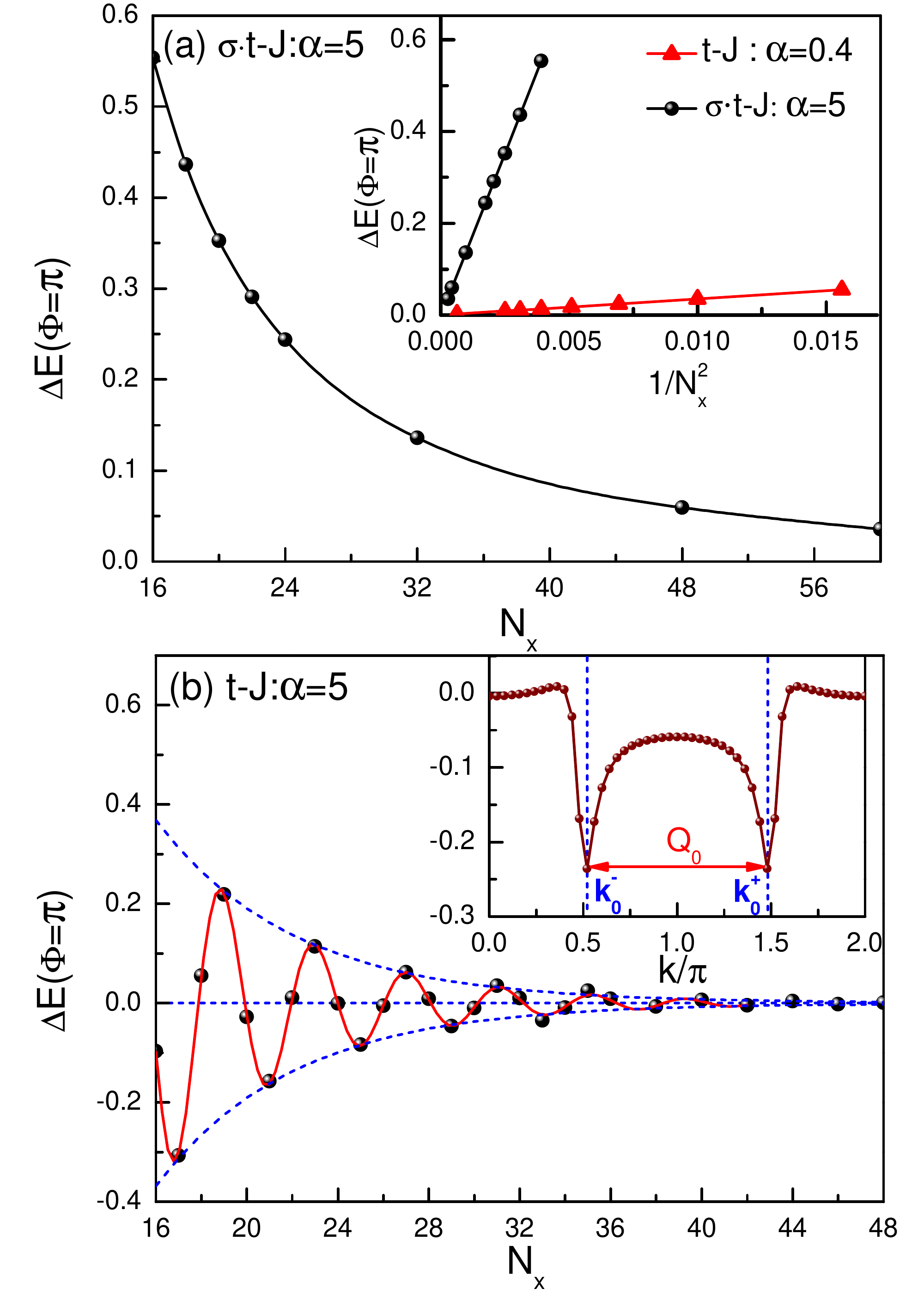}
\end{center}
\par
\renewcommand{\figurename}{Fig.}
\caption{(Color online) The energy change due to the charge response to an inserted flux $\Phi=\pi$ into the ring geometry of the ladder (see text). (a) The typical Bloch wave behavior ($\propto 1/N_x^2$) for the $\sigma$$\cdot$$t$-$J$ case and the $t$-$J$ ladder at $\alpha<\alpha_c$ (the inset, in which the slope for the $t$-$J$ ladder changes sign but not amplitude at $N_x=$ odd);  (b) The non-Bloch-wave response at $\alpha>\alpha_c$ in the $t$-$J$ case. The energy change can be well fitted by Eq.~(\ref{eq:deltaE1}), which directly relates the phase string effect according to Eq.~(\ref{eq:deltaE}) as the underlying cause for charge incoherence and incommensurate momentum splitting as shown in the inset (the Fourier transformation of the energy change vs. $N_x$, which gives rise to the wavevector splitting precisely at $Q_0$ with some intrinsic broadening). }
\label{Fig:inch}
\end{figure}

Define the energy change
\begin{equation}
\Delta E_{G}^{\text{1-hole}} \equiv E_{G}^{\text{1-hole}} (\Phi=\pi) - E_{G}^{\text{1-hole}} (\Phi=0),
\end{equation}
with $\Phi=0$ corresponding to the periodic boundary condition (PBC) and $\Phi=\pi$ the anit-PBC for the hole \cite{ZZ2013,ZZ2014qp}.  Note that  in terms of Eq.~(\ref{Zc}),
$\mathcal{Z}_{t\text{-}J}(\Phi=0) \equiv\sum_{\nu}\mathcal{Z}_{t\text{-}J}^{(\nu)}$ and $\mathcal{Z}_{t\text{-}J}(\Phi=\pi) \equiv\sum_{\nu}(-1)^{\nu}\mathcal{Z}_{t\text{-}J}^{(\nu)} $, with $\mathcal{Z}_{t\text{-}J}^{(\nu)}\equiv\sum_{c_{\nu}}{\tau }_{c_{\nu}}\mathcal{W}[c_{\nu}]$, where  $\nu$ denotes the winding number counting how many times the hole circumvents the ring.
Then, a straightforward manipulation gives rise to
\begin{align}
  \label{eq:deltaE}\nonumber
  \Delta E_{G}^{\text{1-hole}} (t\text{-}J) &= -\lim_{\beta\to\infty} \frac{1}{\beta} \ln \left( \frac{\mathcal{Z}_{t\text{-}J}(\Phi=\pi)}{\mathcal{Z}_{t\text{-}J}(\Phi=0)}  \right)\\
& =2 \sum_{c_1}  \tau_{c_1} \rho_{c_1}  + 2\sum_{c_3} \tau_{c_3} \rho_{c_3}+...  ~,
\end{align}
where the expansions are based on the fact that each term is vanishingly small \cite{Note3} in the large $N_x$ limit, and the weight
\begin{equation}
\rho_{c_{\nu}}\equiv \lim_{\beta\to\infty} \mathcal{W}[c_{\nu}]/(\beta \mathcal{Z}_{t\text{-}J}^{(0)})>0.
\end{equation}

By comparison, for the $\sigma$$\cdot$$t$-$J$ model
\begin{align}
  \label{eq:deltaEsigma}
 \Delta E_{G}^{\text{1-hole}} (\sigma t\text{-}J)  =2 \sum_{c_1}   \rho_{c_1}  + 2\sum_{c_3}  \rho_{c_3}+...
\end{align}
which can be simply obtained by inserting $\tau_{c_{\nu}}=+1$ in Eq. (\ref{eq:deltaE}).

Without the phase string factor $\tau_{c_{\nu}}$, a Bloch-wave behavior of the doped hole is expected as discussed in the previous section. In fact, it has been shown \cite{ZZ2013,ZZ2014qp} that $\Delta E_{G}^{\text{1-hole}} \propto1/(m^*_cN_x^2)$ [cf. the inset of Fig. \ref{Fig:inch}(a), $N_x=\mathrm{even}$ for a bipartite lattice], with the effective mass $m_c^*=m^*_s/q^2$. Here  $m^*_s$ denotes the effective mass obtained \cite{ZZ2013,ZZ2014qp} by the scaling law of the ground state energy under an OBC, and $q$ is the effective charge in units of the bare hole $+e$ as the present energy change is measured by inserting an external flux $\Phi$. One has $m^*_s=m^*_c$ and $q=1$ for a coherent quasiparticle (Bloch wave state) carrying the full momentum, charge and spin of a bare hole. Indeed, as confirmed \cite{ZZ2013,ZZ2014qp} by DMRG, this is true for the $\sigma\cdot$$ t$-$J$ case as well as the $t$-$J$ model at $\alpha<\alpha_c$.
It implies that ${\tau }_{c}$ is either absent (the $\sigma\cdot$$ t$-$J$ model) or gets ``screened out'' (the $t$-$J$ model at $\alpha<\alpha_c$) to play no essential role in $\Delta E_{G}^{\text{1-hole}}$ here.

However, for the $t$-$J$ case at $\alpha>\alpha_c$, a distinct charge response is clearly manifested as shown in Fig.~\ref{Fig:inch}(b) at $\alpha=5$: $\Delta E_{G}^{\text{1-hole}} $ oscillates strongly with $N_x$, which can be fitted by\cite{Note2}
\begin{equation}\label{eq:deltaE1}
\Delta E_{G}^{\text{1-hole}} (t\text{-}J)  \propto \left(\E^{ik_0^+ N_x}+\E^{ik_0^- N_x}\right) g({N_x})~,
\end{equation}
where the incommensurate $k_0^{\pm}$ emerge as indicated in the inset of Fig.~\ref{Fig:inch}(b). A similar behavior has been also shown by DMRG \cite{ZZ2014cm} at $\alpha=1$.

Therefore, Eq.~(\ref{eq:deltaE1}) and Fig.~\ref{Fig:inch}(b) provide a direct measurement of $\sum_{c_1}  \tau_{c_1} \rho_{c_1}$ in Eq.~(\ref{eq:deltaE}) at large $N_x$ (note that $\nu>1$ terms decay faster as $N_x$ increases), in which the incommensurate $Q_0=k_0^{+}-k_0^{-}$ reemerges as the consequence of a weighted average over the phase string factor $\tau_{c_1}$. Here one finds a direct quantitative link between the intrinsic TSB and the microscopic sign structure of the  $t$-$J$ model.

Furthermore, the envelope function, $g(N_x)$, gives rise to the broadening of the peaks $k_0^{\pm}$ as shown by its Fourier transformation in the inset of Fig.~\ref{Fig:inch}(b), which characterizes the incoherent scale for the charge.
In particular, the result implies that $m^*_c \rightarrow \infty$ in the $t\text{-}J$ model, which has been previously interpreted \cite{ZZ2013,ZZ2014qp,ZZ2014cm} as the self-localization of the doped charge at $\alpha>\alpha_c$. Alternatively, such incoherence of the charge, in the energy change in response to a momentum change $\Delta k_0=\Phi/N_x$ upon the inserting flux $\Phi$, can be regarded as $q\rightarrow 0$ (with $m^*_c/m^*_s=1/q^2\rightarrow \infty$) with a finite $m^*_s$ at $\alpha>\alpha_c$ \cite{ZZ2013,ZZ2014qp,WSK2015}. Namely, the doped hole loses its charge and momentum (coherence) to reduce to a neutral \emph{spinon} of mass $m^*_s$ at $\alpha>\alpha_c$, accompanied by an incoherent density modulation.

\section{Discussion}

The main conclusion reached in this work is that a Mott insulator may generally possess a hidden spontaneous translation symmetry breaking in the \emph{charge degree of freedom}. As a specific example involving a two-leg spin ladder, it has been shown above that a doped hole as a ``testing particle'' can manifest such a symmetry breaking by exhibiting a density modulation on top of a smooth charge density background at $\alpha>\alpha_c$. The finite-size analysis of the DMRG results indicates that the corresponding ground state must be composed of two counter-propagating Bloch waves with momenta ${\textbf{k}}_0^{\pm}\neq 0$ superimposed on a TSB component $|\Psi_{\mathrm {inc}}\rangle$ as shown in Eq. (\ref{gs}). Here the presence of $|\Psi_{\mathrm {inc}}\rangle$ is crucial as it does not explicitly satisfy the translational symmetry Eq. (\ref{T}) specified by ${\textbf{k}}_0={\textbf{k}}_0^{\pm}$, such that the ground state cannot be decomposed into a superposition of two degenerate Bloch waves in the thermodynamic limit.

The emergent $|\Psi_{\mathrm {inc}}\rangle$ further illustrates how the \emph{adiabatic continuity hypothesis}, which underlies the validation of a Landau's Fermi liquid theory, breaks down by strong correlation. Note that at $\alpha<\alpha_c$ the hole as a coherent quasiparticle carries the full and quantized (in a finite system) momentum ${\textbf{k}}_0$. Here, the absence of the TSB part in Eq. (\ref{gs}) with $|c_1|^2= 0$ ensures the one-to-one correspondence to protect the hole's coherence as the spin background is fully gapped. But, at $\alpha>\alpha_c$, the hole (charge) has gained the momentum continuum via the incoherent component
$|\Psi_{\mathrm {inc}}\rangle$. In this case, there is no more protection on the coherence of the hole to prevent it from mixing with the incoherent component.

An analytic connection between the incommensurate wave vector $Q_0$ of the density wave and the underlying phase-string sign structure of the doped Mott insulator has been explicitly established. Here, due to the spin gap, the irreparable phase string as the spin defect created by hopping, is picked up entirely by the hole in a form of strongly fluctuating internal $Z_2$ signs. At $\alpha>\alpha_c$, the partial fractionalization of the doped hole results \cite{ZZ2014qp,ZZ2014cm} in an uncompensated phase string effect that generally causes the hole incoherent, besides leading to the total momentum splitting by $Q_0$ and a charge density wave. Since the phase string effect is generically present for a (doped) Mott insulator of any dimensions or doping \cite{Sheng1996,Wu2008sign}, its many-body quantum interference effect \cite{Zaanen2011,Weng2011b}, including the spontaneous translational symmetry breaking of the charge part is thus expected to occur beyond the present two-leg ladder and one-hole case.

Finally, we note that a variational ground state wave function has been recently proposed \cite{Wang2015} for the one-hole-doped $t$-$J$ two-leg ladder, which has the following simple form:
\begin{align}
 \label{eq:wf}
 |\Psi_{\text{G}}\rangle= \sum_{i} {\varphi}_h(i)e^{-i\hat{\Omega}_i} {c}_{i\downarrow}|\phi_0\rangle
\end{align}
where the hole wave function ${\varphi}_h(i)$ is a variational parameter, while the key component is the nonlocal operator $e^{-i\hat{\Omega}_j}$ to keep track of the phase string effect \cite{Wang2015}. Different from a \emph{rigid} ``spin-polaron'' in the Bloch-wave state Eq. (\ref{T}), $e^{-i\hat{\Omega}_j}$ here is nonlocal and only satisfies the many-body translational symmetry involving the \emph{whole} spin background. It has been shown \cite{Wang2015} that $|\Psi_{\text{G}}\rangle$ can be decomposed into a quasi-coherent and an incoherent component as given in Eq. (\ref{gs}). It not only reproduces the QCP $\alpha_c\simeq 0.7$ ($t/J=3$) very accurately, but also consistently predicts the charge density wave and the momentum distribution in excellent agreement with the DMRG results.

\begin{acknowledgements}
Useful discussions with R.-Q. He, H.-C. Jiang, Y. Qi, R.-Y. Sun, J. Zaanen are acknowledged. This work is supported by Natural Science Foundation of China (Grant No. 11534007), MOST of China (Grant No. 2015CB921000, 2017YFA0302902), and US National Science Foundation Grant  DMR-1408560 (D.N.S.).
\end{acknowledgements}

\end{document}